\documentclass[twocolumn,amsmath,amssymb]{revtex4}
\usepackage{dcolumn}% Align table columns on decimal point
\usepackage{bm}% bold math
\usepackage[dvipdfm]{graphicx}
\begin{document}
\newcommand{\obs}[1]{\ensuremath{\overrightarrow{\boldsymbol{#1}}}}
\newcommand{\bol}[1]{\ensuremath{\boldsymbol{#1}}}
\newcommand{\suu}{\ensuremath{|\uparrow\uparrow\rangle}}
\newcommand{\sud}{\ensuremath{|\uparrow\downarrow\rangle}}
\newcommand{\sdu}{\ensuremath{|\downarrow\uparrow\rangle}}
\newcommand{\sdd}{\ensuremath{|\downarrow\downarrow\rangle}}
\newcommand{\suuu}{\ensuremath{|\uparrow\uparrow\uparrow\rangle}}
\newcommand{\suud}{\ensuremath{|\uparrow\uparrow\downarrow\rangle}}
\newcommand{\sudu}{\ensuremath{|\uparrow\downarrow\uparrow\rangle}}
\newcommand{\sudd}{\ensuremath{|\uparrow\downarrow\downarrow\rangle}}
\newcommand{\sduu}{\ensuremath{|\downarrow\uparrow\uparrow\rangle}}
\newcommand{\sdud}{\ensuremath{|\downarrow\uparrow\downarrow\rangle}}
\newcommand{\sddu}{\ensuremath{|\downarrow\downarrow\uparrow\rangle}}
\newcommand{\sddd}{\ensuremath{|\downarrow\downarrow\downarrow\rangle}}
\newcommand{\bsuuu}{\ensuremath{\langle\uparrow\uparrow\uparrow|}}
\newcommand{\bsuud}{\ensuremath{\langle\uparrow\uparrow\downarrow|}}
\newcommand{\bsudu}{\ensuremath{\langle\uparrow\downarrow\uparrow|}}
\newcommand{\bsudd}{\ensuremath{\langle\uparrow\downarrow\downarrow|}}
\newcommand{\bsduu}{\ensuremath{\langle\downarrow\uparrow\uparrow|}}
\newcommand{\bsdud}{\ensuremath{\langle\downarrow\uparrow\downarrow|}}
\newcommand{\bsddu}{\ensuremath{\langle\downarrow\downarrow\uparrow|}}
\newcommand{\bsddd}{\ensuremath{\langle\downarrow\downarrow\downarrow|}}
\newcommand{\balpha}{\boldsymbol{\alpha}}
\newcommand{\bbeta}{\boldsymbol{\beta}}
\newcommand{\bR}{\mathbf{R}}
\newcommand{\br}{\mathbf{r}}
\preprint{APS/123-QED}

\title{A model for conservative chaos\\
constructed from multi-component Bose-Einstein condensates\\
with a trap in 2 dimensions}

\author{Hisatsugu Yamasaki}
\email{hisa@a-phys.eng.osaka-cu.ac.jp}
%\homepage{http://sphinx.a-phys.eng.osaka-cu.ac.jp/}
\affiliation{%
Department of Applied Physics,
Osaka City University\\
Sumiyoshi-ku
Osaka, 558-8585 Japan
}%

\author{Yuhei Natsume}
\affiliation{
SGraduate School of Science and Technology, Chiba University\\
Inage-ku, Chiba, 263-8522 Japan \\
}%

\author{Katsuhiro Nakamura}
\affiliation{%
Department of Applied Physics,
Osaka City University\\
Sumiyoshi-ku
Osaka, 558-8585 Japan
}%

% 
%\affiliation{

\date{\today}% It is always \today, today,

%\maketitle
\begin{abstract}
To show a mechanism leading to the breakdown of a particle picture
for the multi-component Bose-Einstein condensates(BECs)
with a harmonic trap in high dimensions, we investigate
the corresponding 2-$d$ nonlinear Schr{\"o}dinger equation
(Gross-Pitaevskii equation)
with use of a modified variational principle.
A molecule of two identical Gaussian wavepackets has two degrees of
freedom(DFs),
the separation of center-of-masses
and the wavepacket width. Without the inter-component interaction(ICI)
these DFs show independent regular oscillations
with the degenerate eigen-frequencies. The inclusion of
ICI strongly mixes these DFs, generating a fat mode that breaks a
particle picture, which however can be recovered by introducing
a time-periodic ICI with zero average.
In case of the molecule of three wavepackets for a three-component BEC,
the increase of amplitude of ICI
yields a transition from regular to chaotic
oscillations in the wavepacket breathing.
\end{abstract}

\pacs{Valid PACS appear here}% PACS, the Physics and Astronomy
%Use showkeys class option if keyword
\maketitle

%\section{Introduction}
%

Recently a great number of theoretical and experimental
efforts have been devoted to
Bose-Einstein condensates (BECs)\cite{Andre-1,Andre-2}.
As well as single-component BECs, the trapping techniques can
create multi-component condensates which involve inter-component nonlinear
interactions. The multi-component BEC,
far from being a trivial extension of the single-component one, presents
novel and fundamentally different scenarios for its ground state and
excitations. In particular, it has been observed that BEC can reach
an equilibrium state characterized by separation of the components in
different domains\cite{Hall}.

BEC has a dual aspect of waves and particles.
The wave nature is high-lighted in the phenomenon of
interference leading to fringe patterns\cite{Andre-1,Andre-2}.
On the other hand, the particle nature of BECs can be seen in
typical localized states like vortices and solitons. In fact solitons were
observed in the quasi-one dimensional BEC\cite{sol-1,sol-2}.

Among the works that emphasize a role of the particle picture for
BEC {\it with a harmonic trap} in high-dimensions,
those  of P\'erez-Garc\'{\i}a {\it et al.} are the most
noteworthy\cite{Garc0-1,Garc0-2,Garc1,Garc2}.
We focus on their two important assertions.
The first one made for a single-component BEC
is as follows\cite{Garc1}:
If the phase of BEC wave function
will be suitably
corrected,  the center-of-mass
for a wavepacket displaced from the origin
obeys Newtonian dynamics and the
Ehrenfest theorem is valid even for
the nonlinear Schr{\"o}dinger equation (NSE).
Furthermore the center-of-mass is decoupled
from dynamics of the shape of a wavepacket.
The second assertion
is concerned with the
multi-component BEC\cite{Garc2}:
If the distance between wavepackets with each
{\it linked to the individual component} is much
larger than their typical widths, the picture of
soliton molecules composed of solitonic atoms (wavepackets) holds for
the multi-component BEC with a harmonic trap in high-dimensions.
The large distance as above
is guaranteed by a sufficiently large centrifugal force
due to non-vanishing angular momentum.
.

\begin{figure}[htbp]
     \begin{minipage}{.38\textwidth}
      \includegraphics[width=\linewidth]{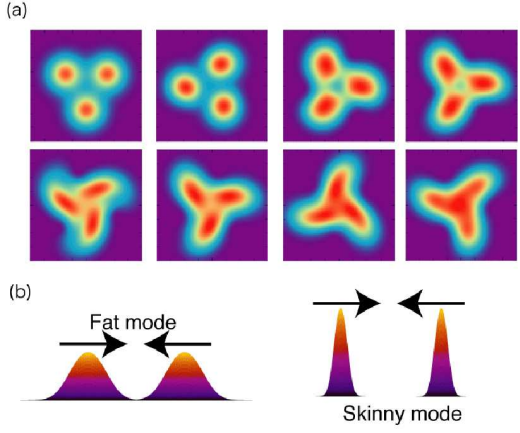}
     \end{minipage}
     \caption{(a) Time evolution (up to $t=30 [\omega^{-1}]$) of initial 
three Gaussian
     wavepackets in 3-component BEC
	with angular momentum $M=9$ and $g=\Lambda g=2\pi$ (computed by using Eq.(\ref{gp})).
	See an outbreak of wave interference;
     (b) Fat and skinny normal modes (schematic).}
\label{fig1}
\end{figure}

P\'erez-Garc\'{\i}a {\it et al.}'s second assertion is
quite interesting because the solitonic structures
like wavepackets are
believed to be dynamically unstable in two and higher spatial dimensions.
However, we have several criticisms
against their assertion:
(1) Inter-component interaction has a tendency
to swell out individual
wavepackets and breaks their picture of interacting particles,
as evidenced in Fig.\ref{fig1}(a);
(2) Collective coordinates
for the width and phase of wave packets,
which should be coupled with the center-of-mass,
are not taken into consideration, although being studied intensively
in other works of their own\cite{Garc0-1,Garc-s3}.

In this Letter,
choosing the $2$-$d$ multi-component BEC
with a harmonic trap, we develop a refined variational principle
to derive an effective dynamics
for interacting wave packets
and pinpoint the weak points
of the idea of P\'erez-Garc\'{\i}a {\it et al.}\cite{Garc2}.
Then we shall propose a new model to restore
the picture of soliton molecules in two dimensions.
In particular, by controlling a parameter
we shall see a transition of the wavepacket breathing
from regular to chaotic oscillations.
We numerically analyze the subject on the
basis of a three-component BEC
corresponding to the ``three-body problem''.

%
%\section{Reduction of the multi-component NSE to effective nonlinear
%dynamics}
The multi-component BEC at zero temperature is
described by NSE (or  Gross-Pitaevskii(GP)
equation).
We shall consider a system of $n$ complex fields
$\psi_1(t,\mathbf{r}),\psi_2(t,\mathbf{r}),\ldots,\psi_n(t,\mathbf{r})$
ruled by the equations
\begin{equation}
i\partial_t \psi_j(t,\mathbf{r}) = \left[
-\frac{1}{2}\Delta_{\mathbf{r}} +
\frac{\mathbf{r}^2}{2}\right]\psi_j(t,\mathbf{r}) +
\sum_{j,k}g_{jk}|\psi_k(t,\mathbf{r})|^2
\psi_j(t,\mathbf{r}) \label{gp}
\end{equation}
for $j,k=1,\ldots,n$ (in units of atomic mass $m=1$, confining length
$\sqrt{\frac{\hbar}{m\omega}}=1$
and oscillation period $\omega^{-1}=1$).
$\mathbf{r}^2/2=(x^2+y^2)/2$ stands for a harmonic trap and
$\sum_kg_{jk}|\psi_k(t,\mathbf{r})|^2$ is the
nonlinear term. The coefficient $g_{ij}$ (proportional to
the scattering length scaled by confining length)
is tunable by Feshbach resonance.
We first consider a static
case of repulsive nonlinearity, $g_{ij}>0$ and choose
$g_{ii}=g$ and $g_{ij}=\Lambda g$ for $i\neq j$.

In the absence of the inter-component interaction(ICI),
i.e., for $\Lambda g=0$,
each component has
a solitonic state, namely Gaussian wavepacket or vortex solution
with a movable center-of-mass.
On the other hand, the non-vanishing ICI($\Lambda g\neq 0$)
among solitons may yield
soliton-soliton interactions.
To reveal a crucial role of the breathing nature of
above localized structures, we apply a refined variational principle
and derive
from (\ref{gp}) the
evolution equation for the collective coordinates of wavepackets.
The trial Gaussian wavepackets(WPs) are constructed from the
circularly-symmetric ground
state solution of (\ref{gp}) with ICI suppressed:
\begin{eqnarray}
\psi_j(\mathbf{r})&=&\sqrt{\frac{1}{\pi w_{j}^2}}
\exp \left[-\frac{(\mathbf{r-R}_j)^2}{2w_{j}^2}
\right]\exp(i\Theta_j(\mathbf{r})),\nonumber\\
\Theta_j(\mathbf{r})&=&\balpha_{j}
(\mathbf{r-R}_j) + \frac{1}{2}(\mathbf{r-R}_j)^{T}
\bbeta_{j}(\mathbf{r-R}_j). \label{wf}
\end{eqnarray}
(One may also choose excited states responsible for vortices.)
${\bf R}=(X,Y)$ is
the center-of-mass and $w$ is width of the circular WP.
\begin{equation}
\balpha=(\alpha^X,\alpha^Y) \label{alpha}
%\mathbf{\alpha}=(\alpha^X,\alpha^Y) \label{alpha}
\end{equation}
    and
\begin{equation}
\bbeta=\left(\begin{array}{cc}
\beta^{XX} &    \beta^{XY}    \\
\beta^{YX}     & \beta^{YY}  \\
\end{array}\right),\qquad \beta^{XY}=\beta^{YX} \label{beta}
\end{equation}
are respectively the first- and second-order coefficients
of Taylor-expansion of the phase $\Theta$ w.r.t.
$\mathbf{r - R}$(a trivial constant phase is suppressed).
${\bf R}$, $w$, $\balpha$ and $\bbeta$ constitute collective coordinates below.

The expansion of $\Theta$ in Eq.(\ref{wf}) is the most natural, although
the existing works\cite{Garc0-1,Garc0-2,Garc-s1,Garc-s3}
employ an expansion
w.r.t. $\mathbf{r}$ rather than $\mathbf{r - R}$.
The advantage of our expansion is that
$\balpha$ and $\bbeta$ turn out to have a transparent correspondence
with the velocity of center-of-mass
and the WP breathing, respectively (see (\ref{ea}) and
(\ref{eb}) below).

According to the variational principle, the action function deriving 
Eq.(\ref{gp})
is obtained from Lagrangian density for
field variables:
\begin{eqnarray}
\mathcal{L} &=& \sum_{j=1}^n  \Bigg[  \frac{i}{2} (\psi_j {\dot
\psi_j^{*}} - \psi_j^{*} {\dot \psi_j}) +
\frac{1}{2} |\nabla_{\mathbf{r}}\psi_j|^2 \nonumber \\
&+& \frac{\mathbf{r}^2}{2}|\psi_j|^2 +
\frac{g}{2} |\psi_j|^4 \Bigg]
+ \sum_{k>j}^{n} \Lambda g |\psi_j|^2|\psi_k|^2 . \label{ld}
\end{eqnarray}
In fact, the multi-component GP equation is obtained from Lagrange
equation,
%\begin{equation}
$\frac{\partial}{\partial t} \frac{\partial \mathcal{L}}{\partial {\dot
\psi_j^{*}}} - \frac{\partial \mathcal{L}}{\partial \psi_j^{*}} +
\nabla \frac{\partial \mathcal{L}}{\partial \nabla \psi_j^{*}} = 0.$
%\label{le}
%\end{equation}

We now insert
(\ref{wf}) into (\ref{ld}) and obtain Lagrangian $L$
for the collective coordinates by integrating
$\mathcal{L}$ over  space coordinates:

%\begin{widetext}
\begin{eqnarray}
L &=& \int \mathcal{L} d\mathbf{r}=
\sum_{j=1}^n \Bigg[  \left( \mathbf{Tr}
{\dot\bbeta}_j + \mathbf{Tr}(\bbeta_{j}^{T}\bbeta_{j})\right)
\frac{w_{j}^2}{4} + \frac{\balpha_{j}^2}{2}\nonumber \\
&-& \balpha_{j} \mathbf{{\dot R}}_j + \frac{{\bf R}_j^2}{2}
+\frac{w_j^2}{2} +
\left(1+\frac{g}{2\pi}\right)\frac{1}{2w_j^2}
\Bigg] \nonumber \\
&+& U(\{\mathbf{R}_j\},\{w_j\})
\label{lg}
\end{eqnarray}
with
\begin{equation}
U(\{\mathbf{R}_j\},\{w_j\})=
\sum_{j>k}^{n}
\frac{\Lambda g}{\pi
(w_j^2+w_k^2)}e^{-\frac{(\mathbf{R}_j-\mathbf{R}_k)^2}{w_j^2+w_k^2}}.
\label{lgg}
\end{equation}
%\end{widetext}
Lagrange equations of motion for phase variables $\balpha$ and
$\bbeta$ lead to
\begin{equation}
      \balpha_j = \mathbf{{\dot R}}_j \label{ea}
\end{equation}
and
\begin{equation}
     \beta_j^{XX}=\beta_j^{YY}= \frac{{\dot w_{j}}}{ w_{j}},
       \qquad \beta_j^{XY}=\beta_j^{YX}=0. \label{eb}
\end{equation}
Noting that $\balpha$ and $\bbeta$ in (\ref{ea}) and (\ref{eb})
(and also all higher-order phase coefficients)
{\it adiabatically follow} ${\bf R}$ and $w$, we
can rewrite Lagrange equations for the latter
by eliminating the former: Equation of motion for ${\bf R}$,
%\begin{equation}
$\frac{d}{dt} \left( \frac{\partial L}{\partial  \mathbf{{\dot R}}_j}
\right) -
\frac{\partial L}{\partial {\bf R}_j} = 0,$
%\label{emr}
%\end{equation}
becomes
\begin{equation}
\mathbf{{\ddot R}}_j +  {\bf R}_j + \frac{\partial }{\partial {\bf R}_j}
U(\{\mathbf{R}_j\},\{w_j\}) = 0. \label{er}
\end{equation}
Similarly for $w$ we have
%\begin{equation}
$\frac{d}{dt} \left( \frac{\partial L}{\partial {\dot w_j}}
\right) - \frac{\partial L}{\partial w_j} = 0,$
%\label{emw}
%\end{equation}
which is reduced to
\begin{eqnarray}
      {\ddot w_j} + w_j - \frac{1}{w_j^3}\left(1+
\frac{g}{2\pi} \right) + \frac{\partial }{\partial w_j}
U(\{\mathbf{R}_j\},\{w_j\})=0.\label{ew}
\end{eqnarray}
%
%\section{2-body problem and stability of WP breathing}

A couple of Eqs. (\ref{er}) and (\ref{ew}) indicate: Contrary to
P\'erez-Garc\'{\i}a {\it et al.}'s theory
that ignored the role of WP width\cite{Garc2},
the width and center-of-mass
are strongly correlated, leading to a breakdown of
the picture of a solitonic molecule
based on the center-of-masses alone. To examine the problem in detail, we
consider the two-component BEC in two dimensions
and explore the fate of a WP breathing against ICI ($\Lambda g$).
Effective Lagrangian leading to Eqs.(\ref{er}) and (\ref{ew}) is now
expressed as
\begin{eqnarray}
L_{\text{eff}} &=&\frac{1}{2}\sum_{j=1}^2\left[  \mathbf{\dot R}_j^2-R_j^2
+ {\dot w_j}^2
- w_j^2 - \frac{1}{w_j^2} \left(1 +
\frac{g}{2\pi}\right)  \right] \nonumber\\
& & - \frac{\Lambda g}{\pi (w_1^2+w_2^2)}
e^{-\frac{(\mathbf{R}_1-\mathbf{R}_2)^2}{w_1^2+w_2^2}} .
\label{Elag}
\end{eqnarray}

Let us define the center-of-mass of two components
${\bf R}_0 = \frac{{\bf R}_1 + {\bf R}_2}{\sqrt{2}}$ and the
relative displacement
$\mathbf{Q} = \frac{{\bf R}_2 - {\bf R}_1}{\sqrt{2}}$,
and suppress
the global translational degree of freedom
(${\bf R}_0=\mathbf{\dot R}_0=0$).
Owing to rotational symmetry
the angular momentum $M$ is a constant of motion:
%\begin{equation}
$M=\frac{\partial L_{eff}}{\partial {\dot \theta}}=
Q^2{\dot\theta}$
%\label{ang}
%\end{equation}
with $\theta$ as a polar angle for $\mathbf{Q}$.
Choosing a synchronous width dynamics $(w_1=w_2=w)$
with the canonical momentum
$p_w=\frac{\partial}{\partial {\dot w}}(L_{eff}-M{\dot \theta})$,
the effective Hamiltonian is given by
\begin{equation}
H_{\text{eff}}=\frac{1}{2}{P_Q}^2
+ \frac{1}{4}{p_w}^2 + V(Q,w)\nonumber\\
\label{nElag}
\end{equation}
with
\begin{eqnarray}
V(Q,w)&=&\frac{1}{2}Q^2+\frac{M^2}{2Q^2}+w^2
    +\frac{1}{w^2}\left(1+\frac{g}{2\pi}\right)
    \nonumber \\
&+& \frac{\Lambda g}{2\pi w^2}e^{-\frac{Q^2}{w^2}}. \label{nEpot}
\end{eqnarray}

Equation (\ref{nEpot}) indicates:
In the absence of ICI($\Lambda=0$), $w$ and $Q$ show
independent regular oscillations
around the minima of individual potentials,
$w=w_0=(1+g/2\pi)^{1/4}$ and $Q=Q_0=\sqrt{M}$, respectively.
Interestingly, however, for any set of values $M$ and $g$ the
typical oscillation frequencies
$\omega_Q^0$ for $Q$ and $\omega_w^0$ for $w$
are always degenerate: $\omega _Q^0=\omega _w^0=2$
(in unit of frequency of the harmonic
potential, $\omega$).
Therefore, by switching on a small
but non-vanishing ICI, dynamics
of $w$ and that of $Q$ will be strongly mixed
to remove the degeneracy and normal
modes, fat (optical) and skinny (acoustic), are formed.
In the fat mode, $Q$ and $w$ change in anti-phase, namely
the decrease of the relative displacement
is accompanied by the blowup of WPs,
while in the skinny mode, $Q$ and $w$ change
in phase (see Fig.\ref{fig1}(b)).
The emergence of the fat mode  brings about
the wave interference between
adjacent WPs as shown in Fig.\ref{fig1}(a).
This provides a mechanism leading to
the breakdown of P\'erez-Garc\'{\i}a {\it et al.}'s
picture of interacting particles.
%
%\section{Time evolution of BEC wave functions under time periodic ICI}
%
\begin{figure}[h]
     \begin{minipage}{.38\textwidth}
      \includegraphics[width=\linewidth]{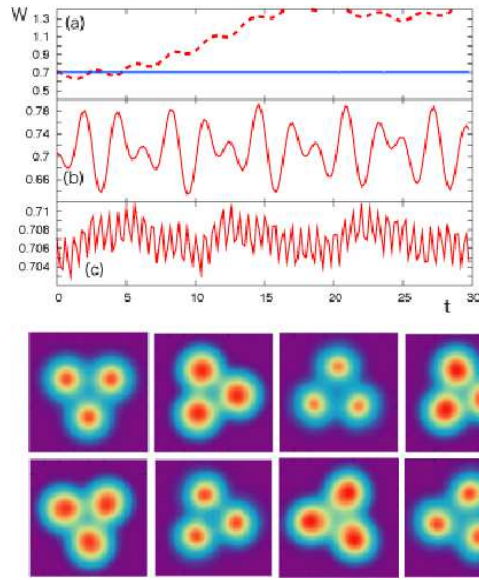}
     \end{minipage}
%\hfill
%   \begin{minipage}{.38\textwidth}
%    \includegraphics[width=\linewidth]{soliton.gif}
%   \end{minipage}
\caption{Upper panel: Time evolution of WP
  width ($w_{j}$). $g=2\pi,\Lambda g=3$
  and angular momentum $M=9$.
  $\Omega$ is a tunable parameter.
(a) $\Omega=0$(dotted line) and $\Omega=100$(solid line),
(b)$\Omega=3$,(c)$\Omega=10$;
Lower panel: Time evolution  (up to $t=100 [\omega^{-1}]$) of initial
three Gaussian WPs in 3-component BEC
with $g=2\pi,\Lambda g=3,\Omega=100$
and $M=9$. See the persistence of a particle picture. }
\label{fig2}
\end{figure}

To suppress the wave interference between adjacent WPs
and recover the particle picture,
we proceed to introduce a time-periodic ICI with zero time-average
and the amplitude $\Lambda g$,
\begin{equation}
g_{ij}=\Lambda g \cos(\Omega t)    \qquad(i\neq j)
\end{equation}
rather than sticking to the static ICI.
Here the frequency scale $\omega_w^0(=2)$
is essential. If $\Omega \gg \omega _w^0$,
the width will feel actually no ICI,
showing a stable and regular breathing proper to a single component BEC
with a harmonic trap. For $\Omega$
near but larger than $\omega_w^0$,
one can expect a sufficiently long stable breathing
until the time ($\gg 2\pi/\omega _w^0$) that a blowup of WP will start.
For $\Omega$ equal or less than $\omega _w^0$,
the breathing oscillation will
quickly blow up leading to the wave interference.

\begin{figure}[t]
     \begin{minipage}{.40\textwidth}
      \includegraphics[width=\linewidth]{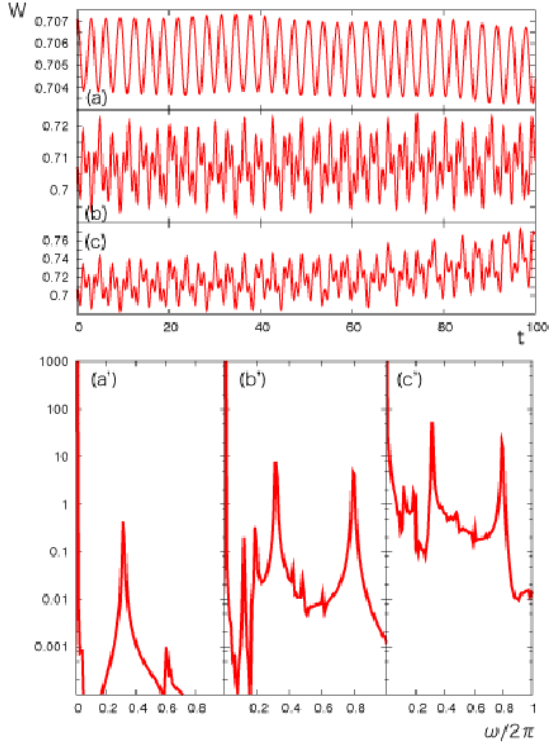}
     \end{minipage}
%\hfill
%     \begin{minipage}{.38\textwidth}
%      \includegraphics[width=\linewidth]{power5.gif}
%     \end{minipage}
\caption{Time evolution of WP width ($w_{j}$) (upper panel) and
corresponding power spectrum (lower panel).
$g=2\pi,\Omega=5$ and $M=9$.
$\Lambda g$ is a tunable parameter. (a)(a')$\Lambda
g=0$,(b)(b')$\Lambda g=2$,
(c)(c')$\Lambda g=4$.}
\label{fig3}
\end{figure}

To verify the above conjecture,
we have numerically integrated 3-component NSE(GP equation)
with a harmonic trap in two dimensions in (\ref{gp})
with use of split-step
Crank-Nicholson method \cite{Mur-1}.
Initial data are coordinates $X_j=R \cos(2 \pi (j-1) / 3),Y_j=R \sin(2 \pi 
(j-1) / 3)$
and velocities $V_{jX}=-R\omega\sin(2 \pi (j-1) / 3),
V_{jY}=R\omega\cos(2 \pi (j-1) / 3)$
with $R^2\omega=M/3$ ($M$: angular
momentum) and $j=1,2,3$. In
short these are placed on the vertices of an equilateral triangle. We have
also calculated expectation values of the widths using $w_{jX}^2=\int
((x-<x>)^2|\psi_j|^2)d\mathbf{r}, w_{jY}^2=
\int ((y-<y>)^2|\psi_j|^2)d\mathbf{r}$ for different times.
In Figs. \ref{fig2} and \ref{fig3} the mean value
$w_j\equiv(w_{jX}+w_{jY})/2$ is plotted as a function of time. Note: even for
a single-component NSE\cite{Garc0-1,Garc0-2}, the initial circular
symmetry of WP is slightly broken during the time evolution, due to
the nonlinear {\it intra-component} interaction; Each of the expectations
$w_{jX}$ and $w_{jY}$ is described by a superposition of modes with nearly-equal frequencies $\omega_w^{0}$ and $\sqrt{(1+g/4\pi)/(1+g/2\pi)}\omega_w^0$
and shows a small beating,
which however can be suppressed by choosing the above mean value $w_j$.

Figure \ref{fig2}(a) shows a blowup of the
breathing oscillation for the static ICI ($\Omega=0$) and
a recovery of stable oscillation
for $\Omega \gg \omega _w^0$, and Figs.\ref{fig2}(b) and \ref{fig2}(c)
show oscillations with no blowup for $\Omega\gtrsim w^0$.
For a fixed value of $\Omega$ in the range $\Omega\simeq\omega _w^0$,
we see a transition from regular
to chaotic breathing oscillations when the amplitude $\Lambda g$ is
increased. In fact, in the case of $\Omega=5$, the oscillation is irregular
for $\Lambda g=2$ and $4$ (see Figs.\ref{fig3}(b) and \ref{fig3}(c)), while
it is regular for $\Lambda g=0$ (see Fig.\ref{fig3}(a)).
The corresponding  power
spectra show a transition from
a distinguished line structure in Fig.\ref{fig3}(a')
to broad ones
in Fig.\ref{fig3}(b')
and \ref{fig3}(c').

We have systematically investigated the wave dynamics
starting commonly from the  equilateral triangle configuration as above.
\begin{figure}[t]
    \begin{minipage}{.38\textwidth}
    \includegraphics[width=\linewidth]{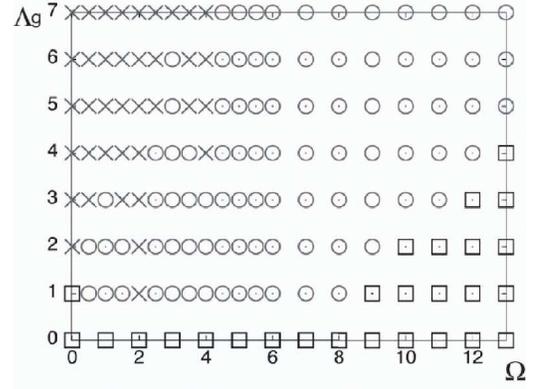}
    \end{minipage}
    \caption{Phase diagram in $\Lambda g$-$\Omega$
     space. ($g=2\pi$ and $M=9$) $\times,\bigcirc,\square$ stands for blowup, chaotic
     and regular regions, respectively. (Data are increased in the left half
 of the diagram for use of detailed discussions.)}
\label{fig4}
\end{figure}
Figure \ref{fig4} is a phase
diagram in $\Lambda g$-$\Omega$ space, which shows three distinct regions,
(i) blowup, (ii) regular without blowup and (iii) chaotic without blowup.
We find the decrease of the blowup region as $\Omega$ is increased beyond 
$\omega_w^0$
and the emergence of chaos without the blowup as $\Lambda g$ is
increased. Contrary to our simple expectations, the phase diagram involves
rich information to be explored: The regular and chaotic oscillations without
blowup are observed even for $\Omega < \omega_w^0$, and
the blowup occurs at resonances,
$\Omega=\omega_w^0$ and $2\omega_w^0$, even for a
small amplitude $\Lambda g$. However, Fig.\ref{fig4}
clearly conveys that the three-component BEC with
the oscillating ICI restores a picture of interacting particles
and can exhibit the transition from regular to
chaotic motions without the blowup. Thus we have obtained a model of 
conservative
chaos constructed from the three-component BEC with a harmonic trap in 2 
dimensions.
%
%\section{Summary and Discussions}
%

To develop P\'erez-Garc\'{\i}a {\it et al.}'s idea,
we have examined
the multi-component repulsive BEC
in a 2-$d$ harmonic trap.
In the absence of the inter-component interaction (ICI), the
wavepacket(WP) breathing and
the motion of relative distance between WPs have
oscillation frequencies always degenerate
for any set of the angular momentum and
intra-component interaction. The non-zero ICI removes the degeneracy
and induces the fat mode that
breaks a picture
of soliton molecules.
We have therefore proposed a new model with a time-periodic
ICI with zero time-average, which can elongate the time interval
without any blowup of WPs.
Provided that the frequency is near the characteristic breathing
frequency ($\omega _w^0$),  a transition from regular
to chaotic oscillations  occurs as the amplitude
of ICI is increased.
For attractive BECs {\it without a trap} or in free space,
the oscillating nonlinearity {\it with non-zero average} is
known to stabilize  high-dimensional
WPs\cite{Saito,Malome}.
In a similar way, in the case of the multi-component repulsive BEC {\it with a trap},
the analogous picture holds  for high-dimensional
WPs  mutually interacting through
time-periodic ICI {\it with zero average}. The amplitude of ICI is found to 
control
the chaoticity of breathing motions of WPs.
The analysis of chaos in space coordinates and the comparison of the present results based on NSE with the effective 
dynamics with a
few degrees-of-freedom in Eqs. (\ref{er}) and (\ref{ew})
will constitute subjects which we intend to study in future.

K.N. is grateful to M. Lakshmanan for valuable conversations
in the early stage of the present work.

%%\bibliography{BEC-Chaos}
\end{document}